\documentclass[cameraready]{Interspeech}

\usepackage[nolist]{acronym}
\usepackage{float}
\usepackage{placeins}
\usepackage{comment}

\begin{acronym}
  \acro{smile}[openSMILE]{Speech \& Music Interpretation by Large-space Extraction}
  \acro{ml}[ML]{Machine Learning}
  \acro{egemaps}[eGeMAPS]{extended Geneva minimalistic acoustic parameter set}
  \acro{qol}[QoL]{Quality of Life}
  \acro{ai}[AI]{Artificial Intelligence}
  \acro{cv}[CV]{Cross-Validation}
  \acro{ci}[CI]{Confidence Interval}
  \acro{ccc}[CCC]{Concordance Correlation Coefficient}
  \acro{pcc}[PCC]{Pearson Correlation Coefficient}
  \acro{mae}[MAE]{Mean Absolute Error}
  \acro{vad}[VAD]{Voice Activity Detection}
  \acro{mnd}[MND]{Motor Neuron Disease}
  \acro{pwmnd}[pwMND]{people with Motor Neuron Disease}
  \acro{als}[ALS]{Amyotrophic Lateral Sclerosis}
  \acro{pma}[PMA]{Primary Muscular Atrophy}
  \acro{sma}[SMA]{Spinal Muscular Atrophy}
  \acro{pwals}[pwALS]{people with ALS}
  \acro{pwsma}[pwSMA]{people with Spinal Muscular Atrophy}
  \acro{pwpma}[pwPMA]{people with Primary Muscular Atrophy}
  \acro{alsfrsr}[ALSFRS-R]{ALS Functional Rating Scale-Revised}
  \acro{qoldys}[QOL-Dys]{Quality of Life in the Dysarthric Speaker}
  \acro{hc}[HC]{Healthy Control}
  \acro{svm}[SVM]{Support Vector Machine}
  \acro{xgb}[XGB]{XGBoost}
  \acro{rf}[RF]{Random Forest}
  \acro{auc}[AUC]{Area Under the Curve}
  \acro{uar}[UAR]{Unweighted Average Recall}
  \acro{asr}[ASR]{Automatic Speech Recognition}
  \acro{loa}[LoA]{Limits of Agreement}
\end{acronym}

\title{Towards Speech Impairment Prediction in German-Speaking Individuals with Amyotrophic Lateral Sclerosis}

\author[affiliation={1, 2, 3}, orcid=0009-0008-9188-058X]{Monica}{Gonzalez-Machorro}
\author[affiliation={4}, orcid=0000-0002-3038-3036]{Ricarda}{von Heynitz}
\author[affiliation={4}, orcid=0009-0000-8371-1278]{Justin}{Hanslmeier}
\author[affiliation={4}]{Finja}{Grimm}
\author[affiliation={1}]{Alexandra-Iulia}{Deac}
\author[affiliation={4}]{Anne}{Gründel}
\author[affiliation={4}, orcid=0000-0002-2078-1997]{Isabell}{Cordts}
\author[affiliation={1,2,3,5},orcid=0000-0002-6478-8699]{Björn}{Schuller}

\address{
    $^1$  CHI -- Chair of Health Informatics, Technical University of Munich University Hospital, Germany \\
    $^2$ audEERING GmbH, Gilching, Germany \\
    $^3$ MCML -- Munich Center for Machine Learning, Germany \\
    $^4$ Department of Neurology, TUM University Hospital – Klinikum Rechts der Isar, TUM School of Medicine and Health, Technical University of Munich, Germany \\
    $^5$ GLAM -- Group on Language, Audio, \& Music, Imperial College London, UK
}

\email{monica.gonzalez@tum.de}


\begin{document}

\maketitle

\begin{abstract}
Amyotrophic Lateral Sclerosis (ALS) is a neurodegenerative disease, often affecting speech due to bulbar dysfunction. In this study, we predict speech impairment in people with ALS (pwALS) using two clinical speech-related scores. We evaluate cross-sectional (across speakers) and personalised (within-speaker) modelling paradigms and analyse the utility of common speech tasks to contribute to the standardisation of speech data collection for pwALS. Experiments on a German-speaking cohort of 66 pwALS show that repetition tasks (/da/-/da/, /da/-/ba/) achieved the best cross-sectional performance (Concordance Correlation Coefficient (CCC) = 0.62) for predicting the Quality of Life in the Dysarthric Speaker questionnaire, while the within-speaker setting reached a CCC of 0.86.
This study represents an initial step towards speech impairment prediction in German-speaking pwALS and highlights the potential of automated speech analysis as a supportive tool for speech impairment assessment.
    
\end{abstract}

\section{Introduction}

\ac{als} is a severe progressive \ac{mnd}, 
with an upper and lower motor neuron pathology leading to reduced mobility, loss of motor control, respiratory failure, and bulbar dysfunction problems~\cite{Foster2019mnd}. 
Median survival of \ac{pwals} is 2-4 years, with respiratory failure as a major cause of mortality~\cite{Feldman2022-amy}. 
%
Since \ac{als} progression is highly heterogeneous, non-linear, and ultimately irreversible~\cite{Turner2013mimics},  
there is a strong need for reliable biomarkers to monitor the disease over time and facilitate intervention~
\cite{Barone2023altered}. 

The \ac{alsfrsr} is the most commonly used clinical measurement for assessing functional impairment in \ac{als}~\cite{Winhammar2005assessment, Cedarbaum1999thealsfrsr}. It consists of 12 items with a maximum score of 48 points. 
It has demonstrated high consistency, reliability, and responsiveness to change~\cite{Winhammar2005assessment}, and serves as a primary outcome measure in clinical trials on disease progression~\cite{Winhammar2005assessment, Stipancic2018minimally}. 
One of its items, \acs{alsfrsr}-speech, evaluates speech 
and is rated using ordinal values from 0 (loss of useful speech) to 4 (normal speech). 
%
Since speech and motor impairments strongly affect \ac{qol}~\cite{Winhammar2005assessment}, a commonly used complementary measure in \ac{pwals} is the 
subscore \textit{Speech Characteristics} of the \ac{qoldys}. This self-administered questionnaire 
provides a subjective assessment of how speech difficulties impact daily life~\cite{Piacentini2011reliability}. 
It contains 10 items rated on a 5-point scale ranging from 0 (never) to 4 (all of the time), assessing the frequency of impaired speech characteristics. The maximum score is 40 points (highest impairment).
Measuring this information is useful for capturing the impact of speech-related therapeutic interventions in the patient's \ac{qol}~\cite{Nogueira2019measuring}.

In recent years, automatic speech analysis has been explored for the recognition and monitoring of \ac{als}~\cite{Bowden2023asystematic}, as speech-based \ac{ml} methods offer a non-invasive and sensitive way to identify and track speech impairments~\cite{Bowden2023asystematic, hecker2022voice}.
Previous works
have reported accuracy scores up to 98\% for distinguishing \ac{hc} from \ac{pwals} with a bulbar onset~\cite{Bowden2023asystematic, ena2022detecting}. Beyond detection, studies have also targeted speech severity: Vieira~et~al.\ achieved a multiclass \ac{auc} of 86\% for predicting the \acs{alsfrsr}-speech score~\cite{Vieira2022machine}, and Mallol-Ragolta~et~al.\ reached an \ac{uar} of 88\% when identifying \ac{pwals} with detectable speech impairments~\cite{Mallolragolta2025early}. Kothare~et~al.\ further modelled individual \ac{alsfrsr} trajectories using acoustic, linguistic, and orofacial cues~\cite{Kothare2025multimodal}. The recent SAND challenge~\cite{SANDchallenge} used an Italian dataset, to tackle (1)~multi-class \ac{als} classification and (2)~\ac{alsfrsr} score prediction. Top participants reached about 60\%~F1 for task~1 and 57\% for task~2, the latter not surpassing the 58\%~baseline~\cite{SANDchallenge}.
%
%


However, current studies show considerably variability in data collection protocols~\cite{Bowden2023asystematic, Qian2025computer}. Although some initial efforts in harmonising data collection for atypical speech have been made~\cite{yue2025challenges}, they are not specific for \ac{als} and still at an early stage. In the case of \ac{als}, standardising speech protocols is particularly important given the burden of the disease. 
Combined with the clinical heterogeneity of \ac{als}~\cite{Turner2013mimics}, these challenges hinder comparison across studies. 
%
In this work, we address these gaps by evaluating: (i) differences in prediction performance between cross-sectional and within-speaker speech impairment models, (ii) differences in predictability between two speech-related clinical scores--the established \ac{alsfrsr}-speech and the more subjective \ac{qoldys} scores-- and (iii) the utility of common speech tasks for predicting speech impairment.
By analysing both scores, multiple speech tasks, and two modelling settings within the same \ac{als} cohort, our study provides a
comparison for speech tasks, modelling settings and clinical scores for speech impairment prediction.
Our research questions are:
\begin{itemize}
\item \textbf{RQ1:} \emph{Prediction of speech impairment scores for \ac{pwals}:} Can acoustic features predict speech impairment both across speakers (cross-sectional) and within individuals during clinical assessments?
\item \textbf{RQ2:} \emph{Comparison of speech impairment scores}: Which of the two speech‑related clinical scores (\ac{alsfrsr}-speech or \ac{qoldys}) is more predictable from acoustic information? 
\item \textbf{RQ3:} \emph{Speech task utility}: Which speech task(s) 
is/are most informative for predicting these scores?
\end{itemize}






The paper is organised as follows. Section~\ref{sec:mandm} presents the dataset and describes the methodology. Sections~\ref{sec:results} and~\ref{sec:discussion} report and discuss the obtained 
results. Finally, Section~\ref{sec:conclusions} concludes the paper and section~\ref{sec:reproducibility} provides information to reproduce the experimental results. 

\section{Material and methods}
\label{sec:mandm}

\subsection{Dataset}
The AIMnd~2.0 dataset is an extension of the dataset 
introduced in~\cite{Mallolragolta2025detection}, collected at the outpatient clinic for \acp{mnd} at the Department of Neurology of the Technical University of Munich University Hospital
in Germany. The data collection received ethical approval (nr.2023-325-S-NP). 
Recording details are available in~\cite{Mallolragolta2025detection}. 

The dataset contains audiovisual recordings of \ac{hc} and \acp{pwmnd}. In this study,
we include only \ac{pwals} (n=66), with up to three sessions per participant. 
We analyse five speech tasks: sustained \textipa{/a:/} (maximum phonation time), the Cookie Theft picture description~\cite{goodglass1983bdae}, two diadochokinetic tasks (/da/-/da/ and /da/-/ba/), and the reading passage ``North Wind and the Sun'' (German version). These tasks are commonly used in speech-based \ac{als} studies~\cite{Bowden2023asystematic, dubbioso2024voice}. 
Other non-speech tasks were also recorded but are out of the scope in this work. 

\subsection{Data processing}

Audio-visual recordings are converted to WAV, downsampled to 16 kHz, and segmented using a \ac{vad} model~\cite{Bredin2021endtoend}\footnote{\url{https://huggingface.co/pyannote/voice-activity-detection}} with default settings; segments shorter than 300 ms were discarded. To reduce background noise--from a laboratory freezer--we apply spectral-gating noise reduction, removing approximately 70\% of stationary noise.

To address RQ1 (\textit{prediction of speech impairment scores for \ac{pwals}}), we define two split strategies: a speaker‑independent split for cross‑sectional modelling and a speaker‑dependent split for within‑speaker setting.
To create the speaker-independent splits (cross-sectional setting), we create three severity bins from the \ac{alsfrsr}-speech score using quantile binning. We perform a speaker-level, stratified split, keeping the distribution of severity bins across training and testing. All sessions from each speaker are kept within the same split to prevent speaker leakage. The test size corresponds to 20\% of the speakers. The resulting cohort, clinical scores, and metadata for each split are reported in Table~\ref{tab:dataset}.
For the within-speaker setting, we define a temporal split. The training set contains all first sessions from each speaker (speakers=66, \ac{alsfrsr}-speech=3.11 $\pm$ 0.99, \ac{qoldys}=14.53 $\pm$ 10.89). The testing set consists of the speakers' second and third sessions, if available (speakers=17, sessions=20, \ac{alsfrsr}-speech=3.10 $\pm$ 0.79, \ac{qoldys}=14.00 $\pm$ 8.78). 
This split ensures that evaluations are performed on future sessions, preventing temporal leakage and mimicking a monitoring scenario in which a clinician uses knowledge from each patient to assess their follow‑up progression.

\begin{table}[t]
\centering
\small
\setlength{\tabcolsep}{2pt}  
\renewcommand{\arraystretch}{0.9}
\caption{Distribution of participants, sessions, demographics, and speech-related clinical scores. Training and testing split distribution for the cross-sectional setting are also available.}
\label{tab:dataset}
\vspace{-1.2em}
\begin{tabular}{p{1.9cm}ccc} 
\toprule
 & \textbf{Total} & \textbf{Training} & \textbf{Testing} \\
\midrule
\textbf{Participants}        & 66  & 52  & 14  \\
\textbf{Sessions}            & 96  & 69  & 27  \\
\textbf{Gender (M/F)}        & 57 / 39 & 42 / 27 & 15 / 12 \\
\textbf{Age}     & 65.99 $\pm$ 11.39 & 66.13 $\pm$ 12.07 & 64.63 $\pm$ 9.65 \\
\textbf{\ac{alsfrsr}-general}    & 33.45 $\pm$ 8.87 & 34.12 $\pm$ 8.09 & 31.74 $\pm$ 10.59 \\
\textbf{\ac{alsfrsr}-speech}            & 3.10 $\pm$ 0.92  & 3.10 $\pm$ 0.96  & 3.11 $\pm$ 0.85 \\
\textbf{QoL-Dys}            & 14.53 $\pm$ 10.36 & 15.09 $\pm$ 10.74 & 13.11 $\pm$ 9.38 \\
\bottomrule
\end{tabular}
\vspace{-1.8em}
\end{table}

\subsection{Feature extraction}

We extract one hand-crafted and two deep-learnt feature sets: 
1) The \ac{egemaps} \cite{Eyben2016thegeneva} from the openSMILE \cite{Eyben2010opensmile} toolkit. This feature set is widely used in similar studies and has shown strong interpretability capacity~\cite{Mallolragolta2025early, xue2019acoustic}. Concretely, we use the 88 functionals and summary statistics from the 25 Low-Level Descriptors (LLDs). These features are also easily reproducible.
2) Embeddings from the transformer‑based Whisper Large v3 model\footnote{\url{openai/Whisper-large-v3}}~\cite{radford2022robust}. We use the 1280‑dimensional encoder representations and, although designed for \ac{asr}, Whisper's internal features capture phonological, linguistic, and prosodic cues across many languages, including German~\cite{george2025advancing}. These properties make the encoder embeddings suited for speech-based prediction tasks, as demonstrated in similar studies~\cite{chaudhari2025self}. 
3) Embeddings from a transformer-based Wav2vec2 model\footnote{\url{facebook/Wav2vec2-large-xlsr-53-german}}~\cite{Conneau21UCR}. These embeddings correspond to the mean pooling of the encoder output and have been used in similar work~\cite{Javanmardi2024pretrained}.
All features were normalised on the training set using standard scaler.  

\subsection{Experimental setup}

\FloatBarrier
\begin{table*}[!h]
\centering
\small
\renewcommand{\arraystretch}{0.6}
\caption{Cross-sectional performance for each speech task for \ac{alsfrsr}-speech and \ac{qoldys} on the testing set. The highest performances are highlighted in bold.}
\label{tab:merged-full}
\vspace{-0.9em}
\begin{tabular}{p{3.4cm} p{1.2cm} p{1.2cm} p{2.4cm} p{1.2cm} p{2.6cm}}
\toprule
\textbf{Task} & \textbf{Model} & \textbf{Features} & \textbf{CCC(CIs)} & \textbf{Bias} & \textbf{$\mathrm{LoA}_{\text{lower}} / \mathrm{LoA}_{\text{upper}}$} \\
\midrule
\multicolumn{6}{l}{\textbf{\ac{alsfrsr}-speech}} \\
\midrule
All tasks fused & SVM & Whisper & 0.56 (0.44–0.66) & 0.11 & -1.06 / 1.30 \\ 
\midrule
/a:/ & SVM & Whisper & 0.45 (0.16–0.63) & -0.10 & -1.50 / 1.31 \\
Picture description & SVM & Whisper & 0.64 (0.45–0.76) & 0.09 & -1.06 / 1.23 \\
/da/–/ba/ & SVM & \ac{egemaps} & 0.50 (0.19–0.68) & 0.19 & -1.10 / 1.48 \\
/da/–/da/ & SVM & \ac{egemaps} & 0.53 (0.31–0.69) & 0.06 & -1.21 / 1.32 \\
Read passage & SVM & Whisper & \textbf{0.65 (0.46–0.76)} & 0.05 & -1.04 / 1.15 \\
\midrule
\multicolumn{6}{l}{\textbf{\ac{qoldys}}} \\
\midrule
All tasks fused & SVM & Whisper & 0.60 (0.45–0.72) & 1.49 & -10.16 / 13.15 \\ 
\midrule
/a:/ & SVM & Whisper & 0.29 (-0.03–0.57) & 2.04 & -15.39 / 19.47 \\
Picture description & XGB & Whisper & 0.50 (0.30–0.65) & 1.86 & -12.10 / 15.81 \\
/da/–/ba/ & SVM & \ac{egemaps} & \textbf{0.62 (0.45–0.72)} & 0.04 & -12.35 / 12.42 \\
/da/–/da/ & SVM & Whisper & \textbf{0.62 (0.39–0.79)} & -0.70 & -14.41 / 13.01 \\
Read passage & SVM & Whisper & 0.54 (0.37–0.69) & 2.55 & -11.97 / 17.07 \\
\bottomrule
\end{tabular}
\vspace{-1.5em}
\end{table*}

\textbf{Regression.}
We employ \acp{svm}, \ac{xgb}, and \acp{rf} for predicting two speech-related clinical scores: the \ac{alsfrsr}-speech and \ac{qoldys} scores. 
According to a systematic review by Bowden~et~al.\ no single model consistently outperforms others in speech-based \ac{als} research; however, \acp{svm} and \acp{rf} generally achieved the strongest performance across prior studies~\cite{Bowden2023asystematic}. 
To address RQ3 (\textit{speech task utility}), each model is trained separately on each speech task.
To obtain a final combined prediction across tasks, we apply mean fusion to the task-specific predictions for each model and feature set.

Model parameters are optimised through a randomised search combined with GroupKFold \ac{cv}, using five folds defined at the speaker-level to prevent speaker leakage between training and validation splits. Experiments are implemented in scikit-learn. Further reproducibility details are available in Section \ref{sec:reproducibility}.~For the \ac{svm} models, we explore the following parameter space:
C~$\in {logspace(-2, 2, 20)}$,
epsilon~$\in {0.05, 0.1, 0.2, 0.5}$,
kernel~$\in {linear, rbf}$,
gamma~$\in {scale, auto}$.
For the \ac{xgb} models, the parameters are:
n\_estimators~$\in {200, 400, 600}$,
learning\_rate~$\in {0.03, 0.05, 0.1}$,
max\_depth~$\in {3, 4, 5, 6}$,
subsample~$\in {0.6, 0.8}$,
colsample\_bytree~$\in {0.6, 0.8}$.
For the \ac{rf} models, we consider:
n\_estimators~$\in {100, 200, 400}$,
max\_depth~$\in {None, 10, 20, 30}$,
min\_samples\_split~$\in {2, 5, 10}$.
Optimal parameters are selected for each model, speech task and feature set. The best-performing configurations are used to retrain the final model, which is then evaluated on the testing set.

\textbf{Evaluation.} Models are evaluated in terms of \ac{ccc} at a speaker- and session-level, meaning we have one inference per session and per speaker.
We also compute the 95\% Confidence Interval (CI) for \ac{ccc}.
The CIs are calculated using 1,000 bootstrapping
iterations~\cite{url2025}. 
To further explore agreement, we calculate Bias (difference between predicted values and ground truth values) and upper and lower \ac{loa}, calculated as 
\[
\text{LoA}_{\text{lower}} = \text{Bias} - 1.96 \, s_d
\]
\[
\text{LoA}_{\text{upper}} = \text{Bias} + 1.96 \, s_d,
\] where \( s_d \) is the standard deviation of the differences, computed as the square root of the sum of squared deviations of each individual difference (\(\hat{y}_i - y_i\)) from the Bias, divided by \(n-1\).


\section{Results}
\label{sec:results}

Tables~\ref{tab:merged-full} and~\ref{tab:combined-rq2} summarise the \ac{ccc} results for predicting the scores \ac{alsfrsr}-speech and \ac{qoldys}. Table~\ref{tab:merged-full} reports cross-sectional performance, whereas Table~\ref{tab:combined-rq2} presents within-speaker \ac{ccc} results. 
For each speech task, only the best-performing model and feature set are shown. Full results and additional metrics are available in the GitHub repository\footnote{\url{https://github.com/monicagoma98/IS_AIMnd_2026}}.
In the cross-sectional setting (Table~\ref{tab:merged-full}), the read passage using Whisper embeddings achieves the highest performance for predicting the \ac{alsfrsr}-speech (\ac{ccc} = 0.65). For \ac{qoldys}, the syllable repetition tasks, /da/-/ba/ and /da/-/da/, using \ac{egemaps} and Whisper embeddings, performs best (\ac{ccc} = 0.62). 
In the within-speaker setting (Table~\ref{tab:combined-rq2}, the read passage again yields the highest performance for \ac{alsfrsr}-speech prediction (\ac{ccc} = 0.71), while the also /da/-/da/ task achieves the strongest results for \ac{qoldys} (\ac{ccc} = 0.86), both with Whisper embeddings.
%
Task fusion for both scores remains lower in both cross-sectional and within-speaker fusion models.

\begin{table}[h!]
\centering
\small
\renewcommand{\arraystretch}{0.9}
\caption{Within-speaker performance for each speech task for \ac{alsfrsr}-speech and \ac{qoldys} on the testing set. The highest performances are highlighted in bold.}
\label{tab:combined-rq2}
\vspace{-0.9em}
\begin{tabular}{p{2.4cm} p{0.9cm} p{1.2cm} p{2.1cm}}
\toprule
\textbf{Task} & \textbf{Model} & \textbf{Features} & \textbf{CCC (CIs)} \\
\midrule

\multicolumn{4}{l}{\textbf{\ac{alsfrsr}-speech}} \\
\midrule
All tasks fused            & \ac{svm} & Whisper     & 0.54~(0.39–0.66) \\
\hline
\textipa{/a:/}             & \ac{rf}  & Whisper     & 0.55~(0.13–0.73) \\
Picture description        & \ac{svm} & Whisper     & 0.62~(0.34–0.74) \\
/da/-/ba/          & \ac{svm} & Wav2vec     & 0.51~(0.11–0.72) \\
/da/-/da/           & \ac{svm} & Whisper     & 0.55~(0.17–0.78) \\
Read passage     & \ac{svm} & Whisper     & \textbf{0.71~(0.55–0.79)} \\
\midrule

\multicolumn{4}{l}{\textbf{\ac{qoldys}}} \\
\midrule
All tasks fused            & \ac{svm} & Whisper     & 0.83~(0.68–0.94) \\
\hline
\textipa{/a:/}             & \ac{xgb} & \ac{egemaps} & 0.51~(0.28–0.67) \\
Picture description        & \ac{svm} & Whisper     & 0.75~(0.54–0.89) \\
/da/-/ba/           & \ac{svm} & Whisper     & 0.79~(0.59–0.90) \\
/da/-/da/           & \ac{svm} & Whisper     & \textbf{0.86~(0.78–0.92)} \\
Read passage     & \ac{svm} & Whisper     & 0.75~(0.54–0.91) \\
\bottomrule
\end{tabular}
\vspace{-1.9em}
\end{table}






\section{Discussion}
\label{sec:discussion}


Regarding RQ1 (\emph{prediction of speech impairment scores for \ac{pwals}}), results on the testing set are promising for most speech tasks and for both settings, with exceptions found mostly for the speech task \textipa{/a:/}. Across tasks and modelling settings, Whisper embeddings combined with \ac{svm} outperform other feature and model combinations in most cases.
As expected, within-speaker predictions (Table \ref{tab:combined-rq2}) outperform cross-speaker results (Table \ref{tab:merged-full}) for both speech impairment scores and for almost all speech tasks, with the exception of picture description and the fused model for \ac{alsfrsr}-speech. In the within-speaker setting, data from the same speakers appear in both training and testing, using a temporal split in which earlier sessions correspond to the training set. The aim of the setup is to account for the heterogeneous progression of \ac{als}. Although simple, our approach demonstrates potential for 
detecting speech impairment in both newly diagnosed German-speaking patients and those undergoing follow-up visits. 
Task fusion shows promising results for \ac{qoldys} in the within-speaker setting. In contrast, fusion-based prediction of the \ac{alsfrsr}-speech score shows slightly higher results for cross-sectional setting over the within-speaker setting, the difference is marginal (0.02) to support solid conclusions.

Addressing RQ2 (\emph{comparison of speech impairment scores}), 
prediction performance was higher for \ac{qoldys} than for \ac{alsfrsr}-speech in both cross-speaker and within-speaker settings. This likely highlights differences in scale design: \ac{qoldys} is a continuous scale from 0 to 40, whereas \ac{alsfrsr}-speech is a coarse 0–4 ordinal scale and is more susceptible to class imbalance with the highest score being overrepresented. 
Further, regression tends to perform better on continuous targets than on ordinal labels and future work should investigate other modelling approaches for ordinal labels \cite{Verhulst2021best}. \ac{qoldys} focuses on speech instability and subtle perceptual changes, which may be earlier reflected in acoustic features, while \ac{alsfrsr}-speech assesses overall functional communication, which tends to be more stable across time and only major changes are then reflected. Also, as the \ac{alsfrsr}-speech was rated by different physicians, there may be inter-rater variability in the \ac{alsfrsr}-speech ratings.
To the best of our knowledge, \ac{qoldys} has not previously been used as a speech-based prediction target. 
We believe future work should combine both measures, as they provide a more comprehensive assessment of speech impairment in \ac{pwals} and may support the development of more effective speech-based \ac{ml} systems for \ac{als}.

Concerning RQ3 (\emph{speech task utility}), the read passage task achieved the highest \ac{ccc} for predicting \ac{alsfrsr}-speech, while /da/-/ba/ and /da/-/da/ performed best for \ac{qoldys} in the cross-sectional setting. Although /da/-/da/ reached a comparatively high \ac{ccc}, it showed underestimation in the predictions (negative mean bias), whereas /da/-/ba/ exhibited minimal bias. 
Across speech tasks, variability in \ac{loa} shows task-dependent prediction stability. For example, the picture description and the read passage tasks show wider \ac{loa} when predicting \ac{qoldys}, indicating high variation of prediction errors despite moderate \ac{ccc} values. 
A similar pattern is observed in the within-speaker setting. For predicting the \ac{alsfrsr}-speech score, the read passage showed the highest agreement, whereas /da/-da/ performed best for \ac{qoldys}. 


The sustained \textipa{/a:/} task exhibited the lowest \ac{ccc} values in both cross-sectional and within-speaker analyses.
Mallol-Ragolta~et~al.\ recently identified \textipa{/a:/} as a highly informative task for distinguishing between \ac{pwals} and \ac{hc} \cite{Mallolragolta2025detection} compared to other commonly used speech tasks. This finding is also supported by other work classifying between \ac{als} and \ac{hc} \cite{vashkevich2021classification}. 
Since our results are inconclusive mostly due to the small sample size, further work is needed to clarify the utility of sustained \textipa{/a:/} in speech impairment prediction in \ac{pwals}.

Diadochokinetic and reading tasks seem more informative for detecting subtle changes in speech impairment, likely because they place greater demands on oral motor control and coordination.~This findings is supported by Mallol-Ragolta~et~al.\,~who found that reading and syllable repetition tasks are among the most salient speech tasks for distinguishing between \ac{hc} and \ac{pwals} with normal speech (\ac{alsfrsr}-speech equal to 4) \cite{Mallolragolta2025early}.
Given that RQ2 suggests that \ac{qoldys} is more sensitive to changes in speech impairment, diadochokinetic tasks, the tasks with the highest \ac{ccc} for this score, may better capture subtle changes in articulation and speech motor coordination.
In contrast, read speech may reflect better functional speech intelligibility, which aligns more to how the \ac{alsfrsr}-speech score is measured~\cite{Winhammar2005assessment}.
In general, our findings suggest that model performance is task-dependent. 
This highlights the importance of understanding speech task utility for speech impairment prediction in \ac{pwals}. 
Current results prevent us from drawing solid conclusion.
The dataset used in this study is derived from a single clinical centre and includes a small cohort. Follow-up data are particularly limited, as only 17 of 66 \ac{pwals} contributed with more than one session.
This highlights the challenges of collecting speech data from \ac{pwals},
as disease progression leads to increased physical impairment and 
potentially 
reduced capacity or willingness to participate. 
Consequently, within-individuals' findings remain preliminary. 

Methodologically, the present analyses are restricted to commonly used and easily reproducible acoustic feature sets. Next steps consist of expanding these to include less frequently explored acoustic features, such as articulatory precision and sentence intelligibility \cite{Talkar2025development, Stegmann2024automated}.
In addition, incorporating linguistic and orofacial features, as proposed in \cite{Kothare2025multimodal}, may further enhance prediction performance of speech impairment. Integrating feature interpretability will also be a key objective in subsequent work.

Future work should also investigate more advanced personalised frameworks, as our current within-speaker approach primarily addresses temporal modelling and does not employ individualised strategies, mostly due to the limited availability of longitudinal samples. 
Further research should also evaluate the generalisability of the proposed approach to related conditions such as other \acp{mnd}, while accounting for potential confounding factors (e.g., medication, weight, \ac{als} onset).
Next steps will also address methodological guidelines specific for the application of \ac{ml} in healthcare such as the CLARITY AI \cite{marconi2024clarity} as well as the clinical usability of speech analysis in healthcare settings.

\section{Conclusion}
\label{sec:conclusions}
In this work, we explored speech impairment prediction in \ac{pwals}. We addressed three research questions: (i) whether acoustic features can predict speech‑related \ac{als} clinical scores both across speakers and within individuals over time; (ii) differences in performance between \ac{alsfrsr}-speech and \ac{qoldys} scores; and (iii) how informative commonly used speech tasks are for this prediction.
Results indicate promising performance
for predicting the two speech-related \ac{als} clinical scores in both a cross-section and within-speaker settings, showing slightly higher performances for predicting the \ac{qoldys} score, suggesting that acoustic features may be more suitable for granular speech changes assessed in this score. Among the five speech tasks, diadochokinetic and the reading tasks yielded the most promising results, whereas the sustained vowel \textipa{/a:/} showed the weakest performance. Future work is needed to draw more solid conclusions.
A system capable of automatically predicting speech impairment in \ac{pwals} could speed up evaluation during clinical assessments. 
Patients could perform speech tasks during waiting times at the clinic or from home. Such system could serve as a support tool for monitoring disease progression.

\section{Reproducibility}
\label{sec:reproducibility}
Unfortunately, due to the ethics approval, the dataset used in this work is not available to the public.~To ensure reproducibility, the underlying code building blocks for this study are available in \url{https://github.com/monicagoma98/IS_AIMnd_2026}.


\section{Acknowledgments}
We thank Adria Mallol-Ragolta for his contributions at the start of the AIMnd project as well as Maxim Korman for his critical intellectual input. We also thank Pascal Hecker and Alexander Gebhard for their valuable feedback while preparing this manuscript. Last but not least, we thank all participants who took part in this study, without your contribution this work could not be possible.



\section{Generative AI Use Disclosure}
This work used Generative AI as a writing assistance tool, specifically for grammar check, improving readability, as well as improving the format of the tables presented. It also assisted code implementations for analyses relevant to this paper.




\end{document}